\documentclass[twocolumn,amsmath,amssymb,aps,footinbib]{revtex4-1}  % 2-column
\usepackage{graphicx}           % for arxiv
\usepackage{dcolumn}
\usepackage{color}
\usepackage{bm}      
\usepackage{xspace}

\newcommand{\be}{\begin{equation}}
\newcommand{\ee}{\end{equation}}
\newcommand{\bea}{\begin{eqnarray}}
\newcommand{\eea}{\end{eqnarray}}
\newcommand{\beaa}{\begin{eqnarray*}}
\newcommand{\eeaa}{\end{eqnarray*}}
\begin{document}
\title{
  General Formula for the Green's Function Approach
  to the Spin-1/2 Antiferromagnetic Heisenberg Model
}
\author{Daiki Sasamoto}
\email{sasamoto.daiki.38n@st.kyoto-u.ac.jp}
\author{Takao Morinari}
 \email{morinari.takao.5s@kyoto-u.ac.jp}
 \affiliation{
  Course of Studies on Materials Science, 
  Graduate School of Human and Environmental Studies, 
  Kyoto University, Kyoto 606-8501, Japan
}
\date{\today}
\begin{abstract}
  A wide range of analytical and numerical methods are available
  to study quantum spin systems.
  However, the complexity of spin correlations and interactions
  limits their applicability to specific temperature ranges.
  The analytical approach utilizing Green's function has proved advantageous,
  as it allows for formulation without restrictions on the presence of long-range order
  and facilitates estimation of the spin excitation spectrum
  and thermodynamic quantities across the entire temperature range.
  In this work, we present a generalized formulation of the Green's function method
  that can be applied to diverse spin systems.
  The notable aspect of our approach is that the theory aligns with
    the high-temperature expansion results in the high-temperature regime. When the spatial
    dimension $d$ is $d\geq 3$, our formulation effectively describes
    the magnetic long-range order as Bose-Einstein condensation occurring
    at the wave vector characteristic of the magnetic long-range order.
    As an application to a frustrated spin system, we consider
    the $J_1$-$J_2$ model on the square lattice, incorporating nematic correlations.
    While our approach may not be robust enough to determine the ground state properties,
    it is capable of describing short-range correlations at finite temperatures
    to a certain extent.
    In the presence of strong frustration,
    accurate numerical calculations become essential.
    Within this framework,
    we compute the specific heat, the spin susceptibility,
    and the spin-wave gaps at characteristic wave vectors,
    and discuss the effects of frustration.
    Our primary focus here is on the spin one-half antiferromagnetic
    Heisenberg model on an arbitrary lattice.
    However, the Green's function approach can be extended to include
    other types of interactions and accommodate higher spin values.
\end{abstract}
\maketitle
%------------------------------------------------------------------------
\section{Introduction}
Quantum antiferromagnetic Heisenberg models are of utmost importance
in exploring intriguing phenomena.
Of particular significance is the case of spin-1/2,
where quantum correlation effects are most pronounced.
A well-known example is cuprate superconductors\cite{Keimer2015},
where high-temperature superconductivity emerges upon hole doping
in the quasi-two-dimensional spin-1/2 Heisenberg antiferromagnet.
The rich physics of cuprate superconductors is likely linked
to the phenomenon of quantum spin liquid.
Originating from Anderson's proposal
of the resonating valence bond state\cite{Anderson1973},
extensive theoretical and experimental research has been devoted
to spin liquid states\cite{Balents2010,Savary2017,Zhou2017},
which may host exotic quantum states.

The exploration of novel quantum ground states poses a challenge
due to their absence of long-range order.
In cases where magnetic long-range order exists, mean-field theory can be applied,
and quantum fluctuations around it can be considered.
However, our particular interest lies in states lacking long-range order
and exhibiting high entanglement.
Analytically investigating such systems is a formidable task,
necessitating the adoption of sophisticated numerical algorithms,
such as the density-matrix renormalization group, tensor network algorithms, and others.

Our interest in quantum spin liquid states extends beyond their ground states.
We must also explore their finite-temperature properties,
especially when considering high-temperature superconductivity
and practical quantum devices.
At finite temperatures, quantum Monte Carlo serves as a powerful numerical algorithm.
However, it may encounter challenges, particularly in cases involving frustration,
where the negative sign problem can arise.

In this paper, we employ a double-time Green's function method\cite{Tyablikov1962,Zubarev1960}.
This approach does not rely on long-range order
  and can be applied to finite-temperature properties.
By employing a decoupling scheme in the equations of motion
that preserves the spin rotational symmetry\cite{Kondo1972},
we derive a general formula independent of spatial dimensionality
and the range of the exchange interaction.
  The notable aspect of our approach is that the theory aligns with
  the high-temperature expansion results in the high-temperature regime.
  When the spatial
  dimension $d$ is $d\geq 3$, our formulation effectively describes
  the magnetic long-range order as Bose-Einstein condensation occurring
  at the wave vector characteristic of the magnetic long-range order.
  As an application to a frustrated spin system, we consider
  the $J_1$-$J_2$ model on the square lattice, incorporating nematic correlations.
  We develop a self-contained formulation of the theory
  that does not require inputs from other methods.
  While our approach may not be sufficiently robust to determine
  the ground state properties,
  it is capable of describing short-range correlations at finite temperatures
  to a certain extent.

The rest of the paper is organized as follows.
  In Sec.~\ref{sec:Formulation}, we derive a general theoretical formula.
  For the case of $d \geq 3$, we present the self-consistent equations,
  including Bose-Einstein condensation, which can occur at the wave vector
  characteristic of the magnetic long-range order.
In Sec.~\ref{sec:d_dim}, we apply this formula to the $d$-dimensional hypercubic lattice.
  We demonstrate that our theory is consistent with other theories
  obtained for $d=1, 2, 3$ at finite temperatures.
In Sec.~\ref{sec:J1J2}, we investigate the $J_1$-$J_2$ model.
We compute the correlation functions at zero temperature.
Despite using only a single parameter for the decoupling approximation,
our results are in good agreement with previous works employing
several decoupling parameters.
  We also explore finite temperature properties to a certain extent.
  We compute the specific heat, the spin susceptibility,
  and the spin-wave gaps at characteristic wave vectors,
  and discuss the effects of frustration.

\section{Formulation}
\label{sec:Formulation}
We consider a general spin-1/2 quantum Heisenberg antiferromagnet
on an arbitrary lattice, with the Hamiltonian expressed as:
\be
H=\sum_{\langle i,j\rangle}J_{ij}\bm{S}_{i}\cdot\bm{S}_{j},
\ee
where $\bm{S}_i = (S_i^x,S_i^y,S_i^z)$ represents the spin-1/2 operator at site $i$.
The parameter $J_{ij}=J_{ji}$ denotes the interaction strength
between the spins at sites $i$ and $j$,
and $\langle i,j\rangle$ indicates the sum over a general pair of spins on the lattice.

In the double-time Green's function method,
we study the equation of motion for the spin Green's function.
To facilitate this analysis, we find it convenient to work with
the Matsubara Green's function rather than the real-time Green's function.
Denoting the imaginary time by $\tau$ and
introducing the imaginary time ordering operator $T_{\tau}$,
the Matsubara Green's function is defined as follows:
\be
\mathcal{G}_{j}(\tau)=-\langle T_{\tau}S_{0}^{+}(\tau)S_{j}^{-}(0)\rangle,
\label{eq:G_spin}
\ee
where $S_j^ \pm  = S_j^x \pm iS_j^y$ and
$A\left( \tau  \right)$ is defined as
$A\left( \tau  \right) = {e^{\tau H}}A{e^{ - \tau H}}$
for any operator $A$.
$0$ represents a site in the bulk of the system.

In order to systematically compute the equation of motion, we first
present a general formula.
Suppose $A$ and $B$ represent the products of spin operators.
We define the following Green's function:
\be
   {{\cal G}_{AB}}(\tau ) =  - \langle {T_\tau }A(\tau )B(0)\rangle
   \equiv {\left\langle {A} | {B} \right\rangle _\tau }.
   \label{eq:G_AB}
   \ee
   The equation of motion is given by
   \be
\frac{\partial }{{\partial \tau }}{{\cal G}_{AB}}(\tau )
=  - \langle {T_\tau }\left[ {H,A(\tau )} \right]B(0)\rangle
- \delta \left( \tau  \right)\left\langle {\left[ {A,B} \right]} \right\rangle.
\label{eq:GAB_EOM}
\ee
The Fourier transform of ${{\cal G}_{AB}}\left( \tau  \right)$ is defined by:
\be
    {{\cal G}_{AB}}\left( {i{\omega _n}} \right)
 = \int_0^\beta  {d\tau } {e^{i{\omega _n}\tau }}{\left\langle {A}
|
 {B} \right\rangle _\tau }
 \equiv {\left\langle {A}
|
 {B} \right\rangle _{i{\omega _n}}}.
 \ee
 Here
 the bosonic Matsubara frequency $\omega_n$ takes the values
 $\omega_n = \frac{2\pi n}{\beta}$, where $n$ is an integer,
 and $\beta = \frac{1}{k_{\rm B} T}$ is the inverse temperature
 with $k_{\rm B}$ being the Boltzmann constant
 and $T$ being the temperature.
 After the Fourier transformation of Eq.~(\ref{eq:GAB_EOM}), we obtain the following equation:
\be
i{\omega _n}{\left\langle {A}
  |
 {B} \right\rangle _{i{\omega _n}}} = {\left\langle {{\left[ {A,H} \right]}}
|
{B} \right\rangle _{i{\omega _n}}} + \left\langle {\left[ {A,B} \right]} \right\rangle.
\label{eq:GAB_formula}
\ee
This is the general equation of motion.
We can use Eq.~(\ref{eq:GAB_formula}) with $A$ replaced by ${\left[ {A,H} \right]}$
to derive the equation of motion for the first term in the right-hand side.
By repeating similar steps, we can derive higher-order equations of motion.
Furthermore, by rewriting Eq.~(\ref{eq:G_AB}) as:
\be
   {{\cal G}_{AB}}(\tau ) =  - \left\langle {{T_\tau }A\left( 0 \right)B\left( { - \tau } \right)} \right\rangle,
   \ee
   we can derive the following equation of motion\cite{Tserkovnikov1981}:
   \be
   i{\omega _n}{\left\langle {A}|
         {B} \right\rangle _{i{\omega _n}}}
   =  - {\left\langle {A}
 \mathrel{\left | {\vphantom {A {\left[ {B,H} \right]}}}
 \right. \kern-\nulldelimiterspace}
         {{\left[ {B,H} \right]}} \right\rangle _{i{\omega _n}}} + \left\langle {\left[ {A,B} \right]} \right\rangle.
   \ee

Applying Eq.~(\ref{eq:GAB_formula}) to Eq.~(\ref{eq:G_spin})
and computing $\left[ {S_0^ + ,H} \right]$
we obtain the following equation of motion:
\be
i\omega_{n}\langle S_{0}^{+}|S_{j}^{-}\rangle_{i\omega_{n}}=J_{\bm{\delta}_{1}}\langle S_{0}^{z}S_{1}^{+}|S_{j}^{-}\rangle_{i\omega_{n}}-J_{\bm{\delta}_{1}}\langle S_{0}^{+}S_{1}^{z}|S_{j}^{-}\rangle_{i\omega_{n}}.
\label{eq:G_spin_EOM1}
\ee
Here, we assume that the SU(2) symmetry in spin space is not broken
so that
\be
\left\langle {\left[ {S_0^ + ,S_j^ - } \right]} \right\rangle  = 2{\delta _{j,0}}\left\langle {S_0^z} \right\rangle  = 0.
\ee
In Eq.~(\ref{eq:G_spin_EOM1}), we denote 1 as the site interacting with site 0,
and $\bm{\delta}_{1}$ represents the displacement vector
connecting site 0 and site 1.
The summation with respect to site 1 is implicit
to simplify the notation.

The equation of motions for the two terms in the right hand side of
Eq.~(\ref{eq:G_spin_EOM1}) are
\begin{widetext}
\bea
i{\omega _n}{\left\langle {{S_0^zS_1^ + }} |
         {{S_j^ - }} \right\rangle _{i{\omega _n}}}
&=&
{J_{{\bm{\delta}_2}}}{\left\langle {{S_0^zS_1^zS_{1 + 2}^ + }} |
 {{S_j^ - }} \right\rangle _{i{\omega _n}}} - {J_{{\bm{\delta}_2}}}{\left\langle {{S_0^zS_1^ + S_{1 + 2}^z}}|
  {{S_j^ - }} \right\rangle _{i{\omega _n}}}
+ \frac{1}{2}{J_{{\bm{\delta}_2}}}{\left\langle {{S_0^ + S_2^ - S_1^ + }} |
  {{S_j^ - }} \right\rangle _{i{\omega _n}}}
\nonumber \\
& & 
- \frac{1}{2}{J_{{\bm{\delta}_2}}}{\left\langle {{S_0^ - S_2^ + S_1^ + }} |
  {{S_j^ - }} \right\rangle _{i{\omega _n}}}
+ 2{{\delta}_{1,j}}\left\langle {S_0^zS_1^z} \right\rangle
- {{\delta}_{0,j}}\left\langle {S_0^ - S_1^ + } \right\rangle,
\eea
and
\bea
i{\omega _n}{\left\langle {{S_0^ + S_1^z}} |
  {{S_j^ - }} \right\rangle _{i{\omega _n}}}
&=& \frac{1}{2}{J_{{\bm{\delta}_2}}}{\left\langle {{S_0^ + S_1^ + S_{1 + 2}^ - }} |
 {{S_j^ - }} \right\rangle _{i{\omega _n}}} - \frac{1}{2}{J_{{\bm{\delta}_2}}}{\left\langle {{S_0^ + S_1^ - S_{1 + 2}^ + }} |
  {{S_j^ - }} \right\rangle _{i{\omega _n}}}
+ {J_{{\bm{\delta}_2}}}{\left\langle {{S_0^zS_2^ + S_1^z}} |
  {{S_j^ - }} \right\rangle _{i{\omega _n}}}
\nonumber \\
& & 
- {J_{{\bm{\delta}_2}}}{\left\langle {{S_0^ + S_2^zS_1^z}} |
  {{S_j^ - }} \right\rangle _{i{\omega _n}}}
- {{\delta}_{1,j}}\left\langle {S_0^ + S_1^ - } \right\rangle
+ 2{{\delta}_{0,j}}\left\langle {S_0^zS_1^z} \right\rangle.
\eea
\end{widetext}
Here, $1+2$ represents the site ${\bm{\delta}}_{1} + {\bm{\delta}}_{2}$.
It is important to note that sites 1 and 2 can be the same site in this expression. 

Now we apply the decoupling approximation
introduced by Kondo and Yamaji\cite{Kondo1972}.
That is, for example,
\be
\langle S_{0}^{z}S_{1}^{z}S_{2}^{+}|S_{j}^{-}\rangle_{i\omega_{n}}\simeq\alpha\langle S_{0}^{z}S_{1}^{z}\rangle\langle S_{2}^{+}|S_{j}^{-}\rangle_{i\omega_{n}},
\label{eq:decoupling}
\ee
where $\alpha$ is a parameter to be determined.
This is understood as follows\cite{Kondo1972}:
\bea
{\left\langle {{S_0^zS_1^zS_2^ + }}|
  {{S_j^ - }} \right\rangle _{i{\omega _n}}}
&=& {\left\langle {{S_0^zS_1^z\left[ {\alpha  + 2\left( {1 - \alpha } \right)S_2^z} \right]S_2^ + }}|
  {{S_j^ - }} \right\rangle _{i{\omega _n}}} \nonumber \\
& \simeq &
\left\langle {S_0^zS_1^z\left[ {\alpha  + 2\left( {1 - \alpha } \right)S_2^z} \right]} \right\rangle {\left\langle {{S_2^ + }} |
  {{S_j^ - }} \right\rangle _{i{\omega _n}}} \nonumber \\
&=& \alpha \left\langle {S_0^zS_1^z} \right\rangle {\left\langle {{S_2^ + }}|
 {{S_j^ - }} \right\rangle _{i{\omega _n}}}.
\eea
\begin{widetext}
After applying this decoupling scheme, we obtain
  \bea
  \lefteqn{
    i{\omega _n}
    {{\cal F}_j}\left( {i{\omega _n}} \right)
  }
  \nonumber \\
  & \simeq &
  \frac{1}{2}\left[ {\left( {1 - {{\delta}_{1,2}}} \right){J_{{\bm{\delta}_2}}}
      \alpha {c_{1-2}}
      + {{\delta}_{1,2}}{J_{{\bm{\delta}_1}}}} \right]{{\cal G}_j}\left( {i{\omega _n}} \right)
  - \frac{1}{2}
  \left[ {\left( {1 - {{\delta}_{1 + 2,0}}} \right){J_{{\bm{\delta}_2}}}
      \alpha {c_{1+2} }
      + {{\delta}_{1 + 2,0}}{J_{{\bm{\delta}_2}}}} \right]
       {{\cal G}_{j - 1}} \left( {i{\omega _n}} \right)
  \nonumber \\
  & & + \frac{1}{2}\left( {1 - {{\delta}_{1 + 2,0}}} \right){J_{{\bm{\delta}_2}}}\alpha
            {c_{1}}{{\cal G}_{j - 1 - 2}\left( {i{\omega _n}} \right)}
            - \frac{1}{2}\left( {1 - {{\delta}_{1,2}}} \right){J_{{\bm{\delta}_2}}}\alpha {c_{1}}{{\cal G}_{j - 2} \left( {i{\omega _n}} \right)}
  \nonumber \\
  & &
  - \frac{1}{2}{{\delta}_{1 + 2,0}}{J_{{\bm{\delta}_2}}}{{\cal F}_j}\left( {i{\omega _n}} \right) + \frac{1}{2}{{\delta}_{1,2}}{J_{{\bm{\delta}_1}}}{{\cal F}_j}\left( {i{\omega _n}} \right) + \left( {{{\delta}_{1,j}} - {{\delta}_{0,j}}} \right){c_{1} },
  \eea
\end{widetext}
where
\be
   {{\cal F}_j}\left( {i{\omega _n}} \right) = {\left\langle {{S_0^zS_1^ + }}
 \mathrel{\left | {\vphantom {{S_0^zS_1^ + } {S_j^ - }}}
 \right. \kern-\nulldelimiterspace}
 {{S_j^ - }} \right\rangle _{i{\omega _n}}} - {\left\langle {{S_0^ + S_1^z}}
 \mathrel{\left | {\vphantom {{S_0^ + S_1^z} {S_j^ - }}}
 \right. \kern-\nulldelimiterspace}
         {{S_j^ - }} \right\rangle _{i{\omega _n}}}.
   \ee
Here, we defined the spin correlation function by the following equation:
\be
c_{i-j}
=c_{j-i}
\equiv2\langle S_{i}^{+}S_{j}^{-}\rangle=4\langle S_{i}^{z}S_{j}^{z}\rangle.
\ee
After the Fourier transformation, 
\be
   {{\cal G}_{\bm{k}}}\left( {i{\omega _n}} \right) = \sum\limits_j {{e^{ - i{\bm{k}}
         \cdot {{\bm{R}}_j}}}{{\cal G}_j}\left( {i{\omega _n}} \right)},
   \ee
\be
   {{\cal F}_{\bm{k}}}\left( {i{\omega _n}} \right) = \sum\limits_j {{e^{ - i{\bm{k}} \cdot {{\bm{R}}_j}}}{{\cal F}_j}\left( {i{\omega _n}} \right)},
   \ee
   where ${\bm{R}}_j$ represents the coordinate of site $j$,
we obtain   
\bea
i{\omega _n}\left( {\begin{array}{*{20}{c}}
{{{\cal G}_{\bm{k}}}\left( {i{\omega _n}} \right)}\\
{{{\cal F}_{\bm{k}}}\left( {i{\omega _n}} \right)}
\end{array}} \right)
& \simeq & \left( {\begin{array}{*{20}{c}}
0&{{J_{{\bm{\delta}_1}}}}\\
{{K_{\bm{k}}}}&0
\end{array}} \right)\left( {\begin{array}{*{20}{c}}
{{{\cal G}_{\bm{k}}}\left( {i{\omega _n}} \right)}\\
{{{\cal F}_{\bm{k}}}\left( {i{\omega _n}} \right)}
\end{array}} \right)
\nonumber \\
& & + \left( {\begin{array}{*{20}{c}}
0\\
{\left( {{e^{i{\bm{k}} \cdot {\bm{\delta}_1}}} - 1} \right){c_{1} }}
\end{array}} \right),
\label{eq:G_F}
\eea
where
\bea
    {K_{\bm{k}}} &=& \frac{1}{2}\left[ {\left( {1 - {{\delta}_{1,2}}} \right){J_{{\bm{\delta}_2}}}\alpha {c_{1-2} }
        + {{\delta}_{1,2}}{J_{{\bm{\delta}_1}}}} \right]
    \nonumber \\
    & &
    - \frac{1}{2}\left[ {\left( {1 - {{\delta}_{1 + 2,0}}} \right){J_{{\bm{\delta}_2}}}\alpha
        {c_{1+2}
        } + {{\delta}_{1 + 2,0}}{J_{{\bm{\delta}_2}}}} \right]{e^{i{\bm{k}} \cdot {\bm{\delta}_1}}}
    \nonumber \\
    & &
    - \frac{1}{2}\left( {1 - {{\delta}_{1,2}}} \right){J_{{\bm{\delta}_2}}}\alpha
    {
      c_{1}
    }
    {e^{i{\bm{k}} \cdot {\bm{\delta}_2}}} \nonumber \\
    & & + \frac{1}{2}\left( {1 - {{\delta}_{1 + 2,0}}} \right){J_{{\bm{\delta}_2}}}\alpha
    {
      c_{1}
    }
    {e^{i{\bm{k}}
        \cdot \left( {{\bm{\delta}_1} + {\bm{\delta}_2}} \right)}}.
    \eea
    By solving Eq.~(\ref{eq:G_F}) with respect to
    ${\cal G}_{\bm{k}} (i\omega_n)$, we obtain
\be
   {\cal G}_{\bm{k}} (i\omega_n) =\frac{J_{\bm{\delta_{1}}}\left(e^{i\bm{k}\cdot\bm{\delta_{1}}}-1\right)c_{1}
   }
   {\left(i\omega_{n}\right)^{2}-\omega_{\bm{k}}^{2}},
\label{eq:G_result}
\ee
where
\bea
\omega_{\bm{k}}^{2}&=&\frac{1}{2}J_{\bm{\delta_{1}}}
\left[(1-\delta_{1,2}) J_{\bm{\delta_{2}}}\alpha
  c_{1-2}
  \right]\nonumber\\
&-&\frac{1}{2}J_{\bm{\delta_{1}}}
\left[\left(1-
  \delta_{1+2,0}
  \right)
  J_{\bm{\delta_{2}}}\alpha
  c_{1+2}
  + \delta_{1+2,0}
  J_{\bm{\delta_{2}}}\right]e^{i\bm{k}\cdot\bm{\delta_{1}}}\nonumber\\
&-&\frac{1}{2}\left(1-
\delta_{1,2}
\right)J_{\bm{\delta_{1}}}J_{\bm{\delta_{2}}}\alpha
c_{1}
e^{i\bm{k}\cdot\bm{\delta_{2}}}\nonumber\\
&+&\frac{1}{2}\left(1-
\delta_{1+2,0}
\right)J_{\bm{\delta_{1}}}J_{\bm{\delta_{2}}}\alpha
c_{1}
e^{i\bm{k}\cdot\left(\bm{\delta_{1}+\bm{\delta}_{2}}\right)}.
\label{eq:w_result}
\eea
Equations~(\ref{eq:G_result}) and (\ref{eq:w_result})
provide us with the general formula.
When applying this formula to a specific system,
we only need to perform the necessary summations with respect
to 1 and 2,
which are implicit in the equations above.
This allows us to efficiently compute the Green's function
and the corresponding frequency for the given spin system
without the need to derive the equations from scratch each time.

The correlation functions $c_{i-j}$
are determined through a self-consistent calculation.
The Fourier transform of ${{{\cal G}_{\bm{k}}}\left( {i{\omega _n}} \right)}$
leads to
\bea
    {{\cal G}_{\bm{k}}}\left( \tau  \right)
    &=& \frac{1}{\beta }\sum\limits_{i{\omega _n}} {{e^{ - i{\omega _n}\tau }}{{\cal G}_{\bm{k}}}\left( {i{\omega _n}} \right)} \nonumber \\
    &=& \frac{{{A_{\bm{k}}}}}{{2{\omega _{\bm{k}}}}}\left( {\frac{{{e^{ - \tau {\omega _{\bm{k}}}}}}}{{{e^{ - \beta {\omega _{\bm{k}}}}} - 1}} - \frac{{{e^{\tau {\omega _{\bm{k}}}}}}}{{{e^{\beta {\omega _{\bm{k}}}}} - 1}}} \right),
   \eea
   where
   \be
      {A_{\bm{k}}} = {J_{{\bm{\delta}_1}}}\left( {{e^{i{\bm{k}} \cdot {\bm{\delta}_1}}} - 1} \right){
        c_{1}
      }
      \ee
The correlation function is given by
\be
   {c_j} = 2\left\langle {S_0^ + S_j^ - } \right\rangle  = \frac{1}{N}\sum\limits_{\bm{k}} {{e^{i{\bm{k}} \cdot {{\bm{R}}_j}}}\frac{{{A_{\bm{k}}}}}{{{\omega _{\bm{k}}}}}\coth \left( {\frac{{\beta {\omega _{\bm{k}}}}}{2}} \right)}.
   \label{eq:c_j}
   \ee
If we introduce a single decoupling parameter $\alpha$ at Eq.~(\ref{eq:decoupling}),
the total number of parameters to be determined,
including $\alpha$, is equal to the number of self-consistent equations that need to be solved.
This can be achieved by utilizing the identity $c_0 = 1$.
However, if additional decoupling parameters are introduced,
we must impose extra conditions or constraints
on the system to guarantee a unique solution.
These additional conditions can be other theoretical results
and/or experimental findings.
We do not introduce additional decoupling parameters,
and employ a single decoupling parameter.
Consequently, the system of equations can be solved independently
without requiring any extra conditions.
This self-contained nature of the formalism streamlines the solution process,
making it more efficient and eliminating the need
for additional constraints or conditions,
which is particularly advantageous in practical applications.

  Now we present formula for thermodynamical quantities.
  The energy of the system is given by
  \be
  E = \sum\limits_{\left\langle {i,j} \right\rangle } {{J_{ij}}\left\langle {{{\bm{S}}_i} \cdot {{\bm{S}}_j}} \right\rangle }  = \frac{3}{4}N\sum\limits_j {{J_{0j}}{c_j}}.
  \ee
  By taking the derivative of $E$ with respect to the temperature, $T$,
  we obtain the specific heat.
  The uniform spin susceptibility is given by
  \bea
  \chi  &=& \frac{{\beta {{\left( {g{\mu _B}} \right)}^2}}}{N}\sum\limits_{i,j} {\left\langle {S_i^zS_j^z} \right\rangle }
  = \frac{{\beta {{\left( {g{\mu _B}} \right)}^2}}}{{2N}}
  \sum\limits_{i,j} {\left\langle {S_i^ + S_j^ - } \right\rangle }
  \nonumber \\
  &=& \frac{{\beta {{\left( {g{\mu _B}} \right)}^2}}}{4}\sum\limits_j {{c_j}}.
  \label{eq:chi_spin}
  \eea
  Here, $g$ is the $g$-factor and $\mu_B$ is the Bohr magneton.
  Note that there is relationship between
  $\langle S_{i}^{+}S_{j}^{-}\rangle$ and $\langle S_{i}^{z}S_{j}^{z}\rangle$
  because of the SU(2) symmetry in spin space.
  The formula for $\chi$ corresponds to the spin susceptibility
  obtained by taking the average of the longitudinal spin susceptibility
  and the transverese spin susceptibility,
  that corresponds to the spin susceptibility of powder samples experimentally.
  Substituting Eq.~(\ref{eq:c_j}) into Eq.~(\ref{eq:chi_spin}), we obtain
  \be
  \chi  = \frac{{\beta {{\left( {g{\mu _B}} \right)}^2}}}{4}\mathop {\lim }\limits_{{\bm{k}} \to 0} \frac{{{A_{\bm{k}}}}}{{{\omega _{\bm{k}}}}}\coth \left( {\frac{{\beta {\omega _{\bm{k}}}}}{2}} \right).
  \ee

  When the spatial dimension $d$ is $d\geq 3$,
  the theory describes the magnetic long-range order at finite temperature.
  Suppose that
  ${{\omega _{\bm{k}}}} = 0$ at ${\bm k} = {\bm Q}$
  when $T \leq T_c$.
  In this case Bose-Einstein condensation occurs at ${\bm k} = {\bm Q}$,
  and Eq.~(\ref{eq:c_j}) is rewritten as
  \bea
      {c_j} &=& 2\left\langle {S_0^ + S_j^ - } \right\rangle
      \nonumber \\
     &=& \frac{1}{N}\sum\limits_{{\bm{k}}\left( { \ne {\bm{Q}}} \right)}
     {{e^{i{\bm{k}} \cdot {{\bm{R}}_j}}}\frac{{{A_{\bm{k}}}}}{{{\omega _{\bm{k}}}}}
       \coth \left( {\frac{{\beta {\omega _{\bm{k}}}}}{2}} \right)}
     + {e^{i{\bm{Q}} \cdot {{\bm{R}}_j}}}{n_{\bm{Q}}},
     \eea
  where $n_{\bm{Q}}$ is associated with the component of Bose-Einstein condensation
  at ${\bm k} = {\bm Q}$.
  Multiplying ${e^{ - i{\bm{Q}} \cdot {{\bm{R}}_j}}}$
  and taking ${\bm{R}}_j \rightarrow \infty$ limit,
  we obtain
  \be
  \mathop {\lim }\limits_{{{\bm{R}}_j} \to \infty }
          {e^{ - i{\bm{Q}} \cdot {{\bm{R}}_j}}}{c_j}
          = 2\mathop {\lim }\limits_{{{\bm{R}}_j} \to \infty }
          {e^{ - i{\bm{Q}} \cdot {{\bm{R}}_j}}}\left\langle {S_0^ + S_j^ - }
          \right\rangle  = {n_{\bm{Q}}}.
          \ee
  This equation implies the presence of the magnetic long-range order
  with the wavelength ${\bm Q}$:
  \be
  \left\langle {{{\bm{S}}_j}} \right\rangle
  = {e^{i{\bm{Q}} \cdot {{\bm{R}}_j}}}{{\bm m}_{\bm{Q}}},
  \ee
  where
  \be
  \left| {{{\bm m}_{\bm{Q}}}} \right| = \sqrt {\frac{3}{4}{n_{\bm{Q}}}}.
  \ee
  Note that there is no specific direction for ${\bm m}_{\bm{Q}}$ in our formulation
  because of SU(2) symmetry in spin space.
  For $T<T_c$ we need to solve the self-consistent equation with including
  $n_{\bm{Q}}$.
  Here, we assumed that the wave vector for the magnetic long-range order
  is specified by the single wave vector ${\bm Q}$, for simplicity.
  It is easy to extend the formula to the magnetic long-ranger oder with
  multiple wave vectors.

\section{$d$-Dimensional Hypercubic Lattice}
\label{sec:d_dim}
In this section, we shall derive the formula for the hypercubic lattice
with a spatial dimension $d$.
  We show that for $d=1$ and $d=2$, our formula yields accurate results
  at high temperatures, and for $d=3$, it predicts the transition temperature
  with greater precision than mean field theory,
  closely aligning with the quantum Monte Carlo results.
We make the assumption that $J_{ij} = J$ if sites $i$ and $j$
are nearest neighbors, and $J_{ij}=0$ otherwise.
By performing the implicit summations in Eq.~(\ref{eq:w_result}), we obtain
\be
\omega _{\bm{k}}^2 = \frac{1}{2}{J^2}z\left( {1 - {\gamma _{\bm{k}}}} \right)\left[ {1 - \left( {1 + z{\gamma _{\bm{k}}}} \right){a_1} + \left( {z - 2} \right){a_{11}} + {a_2}} \right],
\label{eq:omega_k_d}
\ee
where $z=2d$, $a_1 = \alpha c_{1}$ etc., and
\be
   {\gamma _{\bm{k}}} = \frac{1}{d}\sum\limits_{\mu  = 1}^d {\cos {k_\mu }}.
   \ee
Here, we set the lattice constant to unity.
The Green's function is given by
\be
   {{\cal G}_{\bm{k}}}\left( {i{\omega _n}} \right) =  - \frac{{Jz\left( {1 - {\gamma _{\bm{k}}}} \right){c_1}}}{{{{\left( {i{\omega _n}} \right)}^2} - \omega _{\bm{k}}^2}}.
   \ee
   We note that this result includes
   the one-dimensional chain \cite{Kondo1972}
   and the two-dimensional square lattice \cite{Shimahara1991}.

The self-consistent equations are given by
\be
1 =  - \frac{{z{c_1}J}}{N}\sum\limits_{\bm{k}} {\frac{{1 - {\gamma _{\bm{k}}}}}{{{\omega _{\bm{k}}}}}\coth \left( {\frac{{\beta {\omega _{\bm{k}}}}}{2}} \right)},
\label{eq:d:1}
\ee
\be
{c_1} =  - \frac{{z{c_1}J}}{N}\sum\limits_{\bm{k}} {{\gamma _{\bm{k}}}\frac{{1 - {\gamma _{\bm{k}}}}}{{{\omega _{\bm{k}}}}}\coth \left( {\frac{{\beta {\omega _{\bm{k}}}}}{2}} \right)},
\label{eq:d:2}
\ee
\bea
\left( {z - 2} \right){c_{11}} + {c_2}
&=&  - \frac{{z{c_1}J}}{N}\sum\limits_{\bm{k}}
\left( {z\gamma _{\bm{k}}^2 - 1} \right) \nonumber \\
& & \times
     \frac{{1 - {\gamma _{\bm{k}}}}}{{{\omega _{\bm{k}}}}}
     \coth \left( {\frac{{\beta {\omega _{\bm{k}}}}}{2}} \right).
\label{eq:d:3}
\eea
  Here, $c_{11}$ represents the correlation function between sites
  with a real space displacement vector of $(\pm 1,\pm 1)$
  for $d=2$
  and $(\pm 1,\pm 1, 0)$, $(\pm 1, 0, \pm 1)$, and $(0, \pm 1,\pm 1)$
  for $d=3$.
  For $d=1$, the left hand side of Eq.~(\ref{eq:d:3})
  is reduced to $c_2$ and concurrently, 
  the coefficient of $a_{11}$ vanishes
  in the right hand side of Eq.~(\ref{eq:omega_k_d})
  due to $z=2$.
  Therefore, $c_{11}$ is irrelevant for $d=1$.

\begin{figure}[htbp]
  \includegraphics[width=0.9 \linewidth, angle=0]{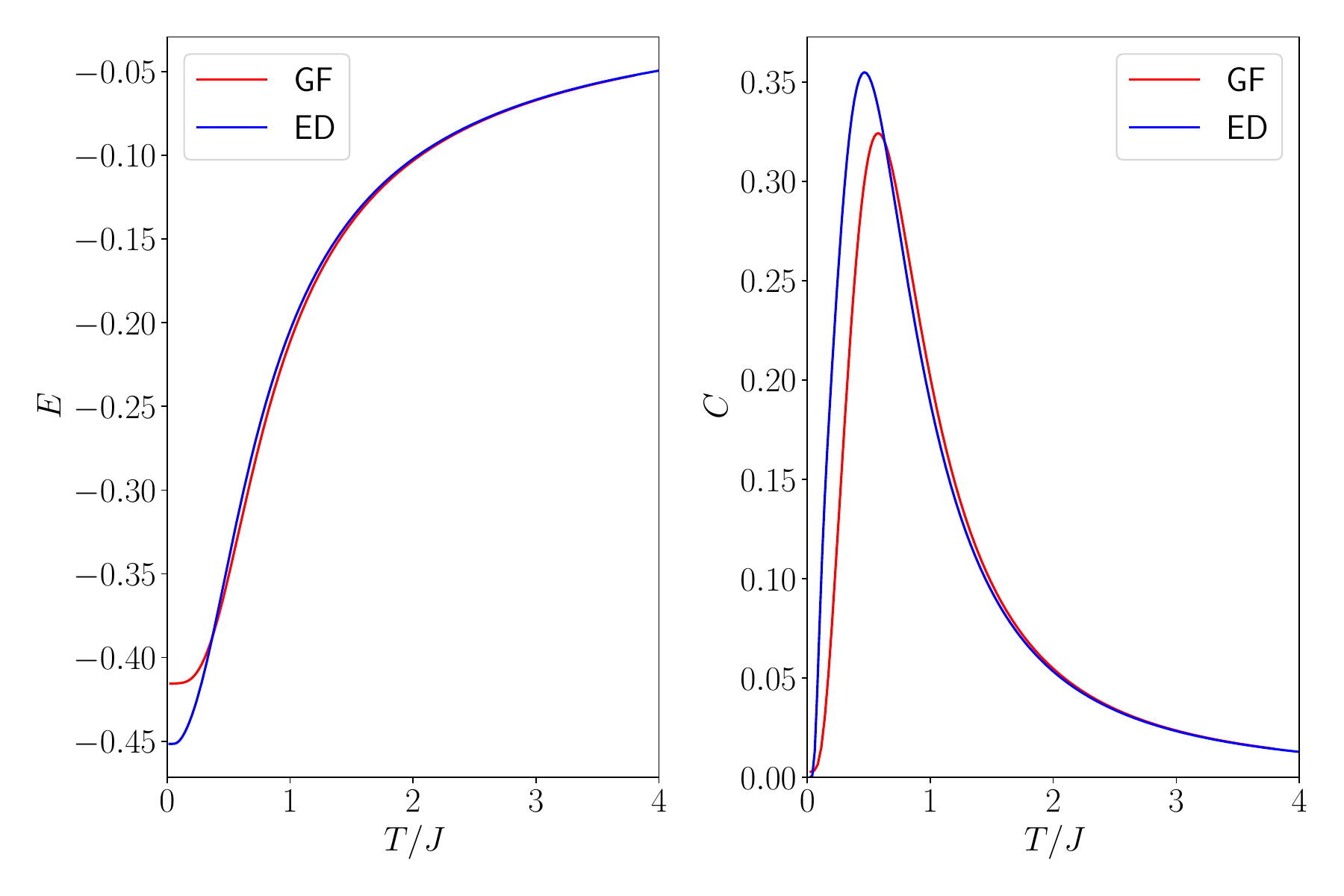}
  \caption{
    \label{fig:1d:E_C}
    (Color online)
      Comparison of the Green's function theory (GF) for the case of $d=1$ with
      the exact diagonalization result (ED)
      for a system of 10 spins: the left panel displays the energy of the system,
      while the right panel presents the specific heat.
  }
\end{figure}

  Figure \ref{fig:1d:E_C} shows the temperature dependence
  of the energy and specific heat for the case of $d=1$.
  This result is benchmarked against the exact diagonalization result
  for a system of 10 spins.
  The Green's function approach aligns exceptionally well
  with the exact diagonalization result at high temperatures.
  However, it deviates from the exact result as the temperature lowers.
  The specific heat peak is overestimated by approximately 16 percent
  compated with the transfer-matrix renormalization-group method\cite{Xiang1998}.
  Detailed discussions comparing the theory with high-temperature expansions
  and the results for the ferromagnetic case are provided
  in ref.~\onlinecite{Kondo1972}.

%-------------------------------------------------------------------
%-------------------------------------------------------------------

\begin{figure}[htbp]
  \includegraphics[width=0.8 \linewidth, angle=0]{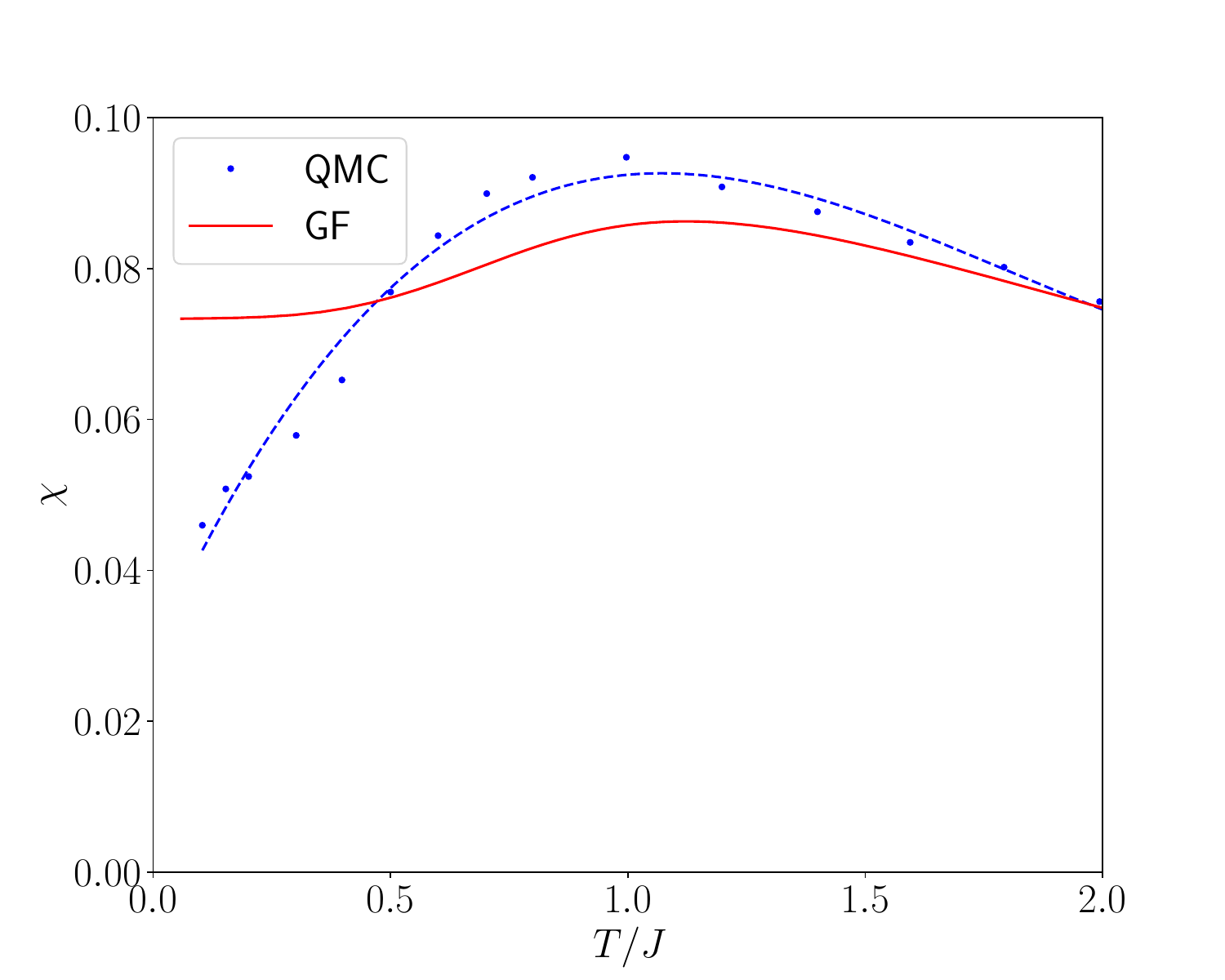}
  \caption{
    \label{fig:2d:chi}
    (Color online)
    Temperature dependence of the spin susceptibility,
    measured in units of ${\left( {g{\mu _{\rm{B}}}} \right)^2}/J$,
      predicted by the Green's function theory (GF) for the case of $d=2$,
      compared with the quantum Monte Carlo results (QMC)\cite{Okabe1988}.
      For the latter, the linear size of the system is $L=12$,
      which is consistent with results obtained for larger $L$\cite{Kim1998}.
      The dashed line is a guide for the eyes.
  }
\end{figure}

  Figure \ref{fig:2d:chi} displays the temperature dependence
  of the spin susceptibility for the case of $d=2$.
  The Green's function theory is benchmarked against
  the quantum Monte Carlo results\cite{Okabe1988,Kim1998}.
  At high temperatures, the Green's function theory closely matches
  the quantum Monte Carlo results, particularly for the peak position
  of the susceptibility.
  However, notable discrepancies arise at low temperatures.
  Methods to improve these results are discussed in ref.~\onlinecite{Shimahara1991},
  which includes a discussion on Bose-Einstein condensation at zero temperature.
  Leveraging the accurate peak position estimation
  provided by the Green's function approach,
  the magnetic torque experiments
  in cuprate high-temperature superconductors
  are satisfactorily analyzed in ref.~\onlinecite{Morinari2018}.

%-------------------------------------------------------------------
%-------------------------------------------------------------------

  We now consider the case of $d=3$.
  In this case, the Green's function theory accounts for the antiferromagnetic
  phase transition at finite temperature.
Antiferromagnetic long-range order occurs
when $\omega_{\mathbf{Q}}$ vanishes,
where $\mathbf{Q} = (\pi, \pi, \pi)$.

%-------------------------------------------------------------------
\begin{figure}[htbp]
  \includegraphics[width=1 \linewidth, angle=0]{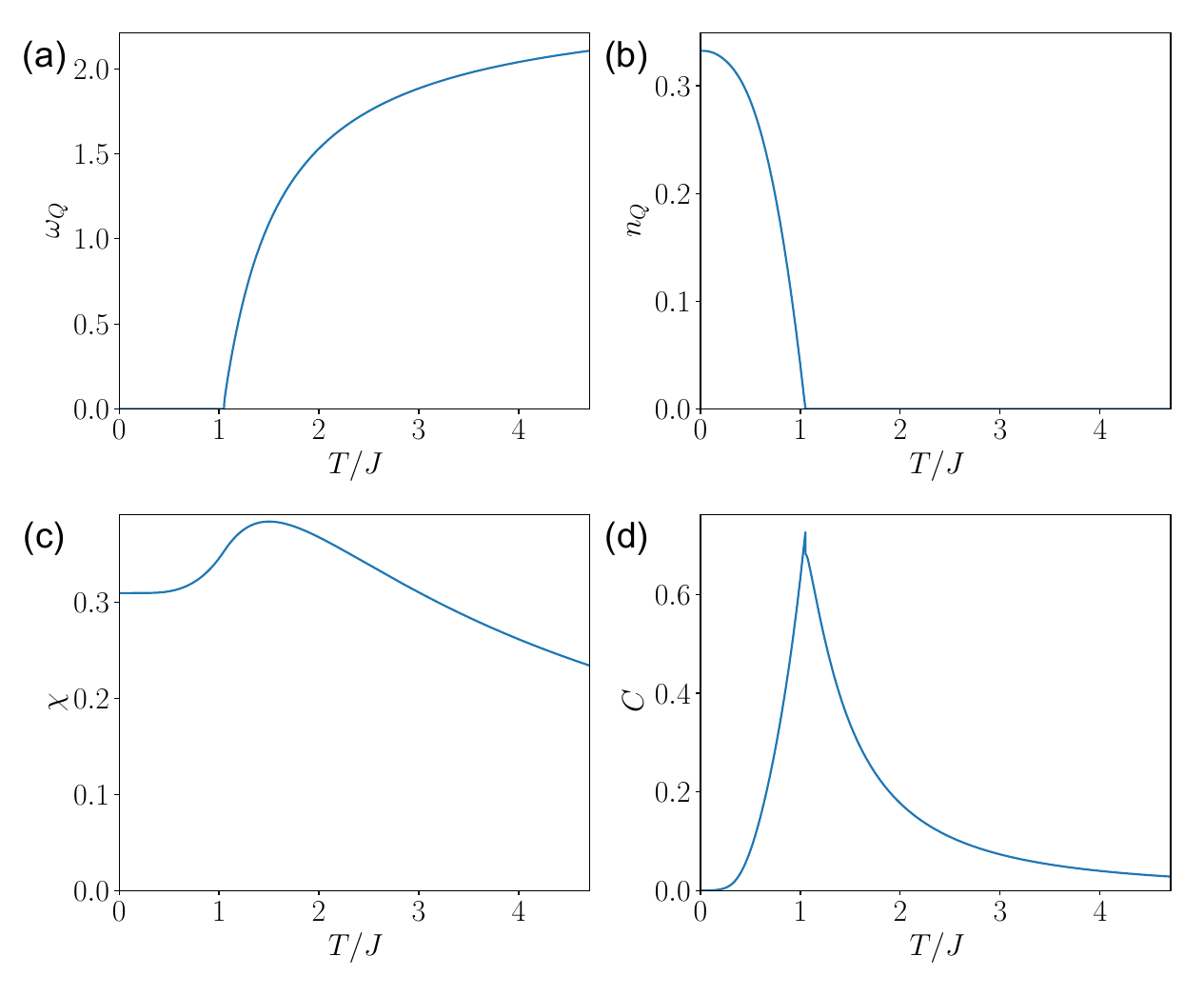}
  \caption{
    \label{fig:3d}
    (Color online)
      Temperature dependencies for the case of $d=3$ showing
      (a) the gap $\omega_{\bm Q}$,
      (b) the Bose-Einstein condensate component $n_{\bm Q}$,
      (c) the spin susceptibility,
      and (d) the specific heat.
      Here, $\omega_Q$ is measured in units of $J$,
      $\chi$ in units of $(g\mu_B)^2/(6J)$, and $C$ in units of $Nk_{\rm B}J$.
  }
\end{figure}
%-------------------------------------------------------------------

  Figure \ref{fig:3d} presents the temperature dependencies
  of the gap $\omega_{\bm Q}$,
  the Bose-Einstein condensate component $n_{\bm Q}$,
  the spin susceptibility,
  and the specific heat.
  To estimate the transition temperature,
  we performed calculations for various values
  of the number of points in the Brillouin zone, $N_{\rm BZ} \equiv N_k^3$.
  Fitting the results to the function $T_c (N_k) = a + b/N_k + c/N_k^2$
  with $a$, $b$, and $c$ as fitting parameters,
  we obtained $T_c = 1.039$,
  which agrees with the previous result\cite{Barabanov1994}.
  This value closely matches the quantum Monte Carlo result\cite{Sandvik1998}
  of $T_c/J = 0.946$.
  Since SU(2) symmetry in spin space is preserved in the theory,
  the spin susceptibility is the average of
  the longitudinal and transverse spin susceptibilities
  as stated above.
  The specific heat exhibits a peak at the transition temperature.

\section{The $J_1$-$J_2$ model}
\label{sec:J1J2}
In the investigation of quantum spin liquid states,
the influence of frustration effects is significant.
To assess the viability of the Green's function approach
for addressing frustrated spin systems,
we turn our attention to the $J_1$-$J_2$ model on the square lattice\cite{Chandra1988,Gelfand1989,Dagotto1989,Figueirido1990,Jiang2012,Hu2013,Gong2014,Wang2018,Ferrari2020,Nomura2021}.
In this model, $J_1$ denotes the interaction between nearest neighbors,
while $J_2$ denotes the interaction between next-nearest neighbors.
%---------------------------------------------------------------------------
This system demonstrates characteristics of frustration,
and it has been hypothesized that a quantum spin liquid phase emerges
around the ratio $J_2/J_1 = 1/2 $ \cite{Chandra1988}.
Recent studies employing machine learning methodologies\cite{Nomura2021}
have suggested that this spin liquid state is realized
within the range of $0.49 \leq J_2/J_1 \leq 0.54 $. 
Meanwhile, investigations utilizing tensor network approaches
has indicated\cite{Gauthe2022} potential
$C_{4v}$ symmetry breaking at $T/J_1 \sim 0.1$ when $J_2 \sim J_1$.

  In the following, we investigate this model at both zero and finite temperatures.
  Although a discussion of the conditions for the spin liquid state is precluded
  due to the limitations of the second-order approximation used in this study,
  we can nonetheless gain insights into $C_{4v}$ symmetry breaking
  or nematic correlations by considering the anisotropy of spin correlations.
%---------------------------------------------------------------------------

By employing the derived formula above, we obtain the following results:
\begin{widetext}
\bea
\omega _{\bm{k}}^2
&=& 2J_1^2\left[ {\left( {1 + {a_2} + 2{a_{11}}} \right) - \left( {1 + 4{\gamma _{\bm{k}}}} \right){a_{1s}}} \right]\left( {1 - {\gamma _{\bm{k}}}} \right) + 2J_1^2\gamma _{\bm{k}}^a\left( {1 + 4{\gamma _{\bm{k}}}} \right){a_{1a}} \nonumber \\
& & + 2J_2^2\left( {1 - {\gamma^{\prime}_{\bm{k}}}} \right)\left[ {\left( {1 + {a_{22}} + 2{a_2}} \right) - {a_{11}}
    \left( {1 + 4 {\gamma^{\prime}_{\bm{k}}}  } \right)
  } \right] \nonumber \\
& & + 4{J_1}{J_2}\left[ {\left( {1 - {\gamma _{\bm{k}}}} \right)
    + \left( {1 -
      {\gamma^{\prime}_{\bm{k}}}
    } \right) - 2
    {\gamma^{\prime}_{\bm{k}}}
    \left( {1 - {\gamma _{\bm{k}}}} \right)} \right]{a_{1s}} + 4{J_1}{J_2}\gamma _{\bm{k}}^a\left( {1 + 2
  {\gamma^{\prime}_{\bm{k}}}
} \right){a_{1a}} \nonumber \\
& &
+ 4{J_1}{J_2}\left[ {\left( {1 - {\gamma _{\bm{k}}}} \right)
    + \left( {1 -
      {\gamma^{\prime}_{\bm{k}}}
    } \right)} \right]{a_{21s}} - 4{J_1}{J_2}\gamma _{\bm{k}}^a{a_{21a}} - 8{J_1}{J_2}{\gamma _{\bm{k}}}\left( {1 -
  {\gamma^{\prime}_{\bm{k}}}
} \right){a_{11}},
\label{eq_wJ1J2}
\eea
\end{widetext}
where
${\gamma _{\bm{k}}} =\left( {\cos {k_x} + \cos {k_y}} \right)/2$,
\be
   {\gamma^{\prime}_{\bm{k}}}
   = \cos {k_x}\cos {k_y},
   \label{eq:gamma_kp}   
   \ee
   \be
   \gamma _{\bm{k}}^a = \frac{1}{2}\left( {\cos {k_x} - \cos {k_y}} \right),
   \label{eq:gamma_ka}
\ee
and
${a_{1s}} = \left( {{a_{1x}} + {a_{1y}}} \right)/2$,
${a_{1a}} = \left( {{a_{1x}} - {a_{1y}}} \right)/2$,
${a_{21s}} = \left( {{a_{21x}} + {a_{21y}}} \right)/2$,
${a_{21a}} = \left( {{a_{21x}} - {a_{21y}}} \right)/2$.
$a_{1x}$ ($a_{1y}$) refers to the nearest neighbor correlation function
in the $x$ ($y$) direction.
Similarly, $a_{21x}$ ($a_{21y}$) denotes the correlation
function between the sites separated by the displacement vector $(2,1)$ ($(1,2)$).
  The variables $a_{1a}$ and $a_{21a}$ are associated with
  $C_{4v}$ symmetry breaking.
When $a_{1a}$ and $a_{21a}$ are non-vanishing,
that indicates the presence of nematic correlations in the system.
We note that Eq.~(\ref{eq_wJ1J2}),
when excluding these nematic correlations,
agrees with the previous result\cite{Siurakshina2001}.
The self-consistent equations are given by
\bea
\alpha  &=& {a_{1s}}I_{00}^{\left( 0 \right)} + {a_{1a}}I_{00}^{\left( 1 \right)} + {a_{11}}I_{00}^{\left( 2 \right)},
\label{eq:J1J2sceq_start}\\
{a_{1s}} &=& {a_{1s}}I_{1s}^{\left( 0 \right)} + {a_{1a}}I_{1s}^{\left( 1 \right)} + {a_{11}}I_{1s}^{\left( 2 \right)},\\
{a_{1a}} &=& {a_{1s}}I_{1a}^{\left( 0 \right)} + {a_{1a}}I_{1a}^{\left( 1 \right)} + {a_{11}}I_{1a}^{\left( 2 \right)},\\
{a_{11}} &=& {a_{1s}}I_{11}^{\left( 0 \right)} + {a_{1a}}I_{11}^{\left( 1 \right)} + {a_{11}}I_{11}^{\left( 2 \right)},\\
{a_2} &=& {a_{1s}}I_{20}^{\left( 0 \right)} + {a_{1a}}I_{20}^{\left( 1 \right)} + {a_{11}}I_{20}^{\left( 2 \right)},\\
{a_{21s}} &=& {a_{1s}}I_{21s}^{\left( 0 \right)} + {a_{1a}}I_{21s}^{\left( 1 \right)} + {a_{11}}I_{21s}^{\left( 2 \right)},\\
{a_{21a}} &=& {a_{1s}}I_{21a}^{\left( 0 \right)} + {a_{1a}}I_{21a}^{\left( 1 \right)} + {a_{11}}I_{21a}^{\left( 2 \right)},\\
{a_{22}} &=& {a_{1s}}I_{22}^{\left( 0 \right)} + {a_{1a}}I_{22}^{\left( 1 \right)} + {a_{11}}I_{22}^{\left( 2 \right)},
\label{eq:J1J2sceq_end}
\eea
where
\be
I_\eta ^{\left( \ell  \right)} =  - \frac{{8{J^{\left( \ell  \right)}}}}{{\beta N}}\sum\limits_{\bm{k}} {{f_\eta }\left( {\bm{k}} \right)\sigma _{\bm{k}}^{\left( \ell  \right)}g\left( {\frac{{\beta {\omega _{\bm{k}}}}}{2}} \right)},
\label{eq:I}
\ee
with ${J^{\left( 0 \right)}} = {J^{\left( 1 \right)}} = {J_1}$,
${J^{\left( 2 \right)}} = {J_2}$,
$g(x)=x \coth x$, 
and
\bea
\sigma _{\bm{k}}^{\left( 0 \right)}
&=& \frac{{1 - {\gamma _{\bm{k}}}}}{{\omega _{\bm{k}}^2}},\\
\sigma _{\bm{k}}^{\left( 1 \right)}
&=& - \frac{{\gamma _{\bm{k}}^a}}{{\omega _{\bm{k}}^2}},\\
\sigma _{\bm{k}}^{\left( 2 \right)}
&=& \frac{{1 -
    {\gamma^{\prime}_{\bm{k}}}
}}{{\omega _{\bm{k}}^2}}.
\eea
The function ${f_\eta }\left( {\bm{k}} \right)$ are defined by
\bea
   {f_{00}}\left( {\bm{k}} \right) &=& 1,\\
   {f_{1s}}\left( {\bm{k}} \right) &=& {\gamma _{\bm{k}}},\\
   {f_{1a}}\left( {\bm{k}} \right) &=& \gamma _{\bm{k}}^a,\\
   {f_{11}}\left( {\bm{k}} \right) &=&
   {\gamma^{\prime}_{\bm{k}}},\\
   {f_{20}}\left( {\bm{k}} \right) &=& 4\gamma _{\bm{k}}^2
   - 2{\gamma^{\prime}_{\bm{k}}} - 1,\\
   {f_{21s}}\left( {\bm{k}} \right) &=&
   {\gamma _{\bm{k}}}\left( {2
     {\gamma^{\prime}_{\bm{k}}}
     - 1} \right),\\
   {f_{21a}}\left( {\bm{k}} \right) &=&
   \gamma _{\bm{k}}^a\left( {2
     {\gamma^{\prime}_{\bm{k}}}
     + 1} \right),\\
          {f_{22}}\left( {\bm{k}} \right) &=&
          4{\gamma^{\prime}_{\bm{k}}}^2 - 8\gamma _{\bm{k}}^2 + 4{\gamma^{\prime}_{\bm{k}}}+ 1.
\eea
We solve the self-consistent equations (\ref{eq:J1J2sceq_start})
to (\ref{eq:J1J2sceq_end}) as follows:
Initially, we solve the equations for $r=0$,
resulting in a reduced set of equations that simplifies to a single equation.
This equation can be solved using the bisection method.
Subsequently, we solve the differential equations for $a_{ij}$,
employing either the parameter $r$ or the temperature $T$ as the variable.
Notably, it is unnecessary to compute $\alpha$ at intermediary steps.
The change in $\alpha$ concerning $r$ or $T$ can be calculated
at the end of the computation.

\begin{figure}[htbp]
  \includegraphics[width= 0.9\linewidth]{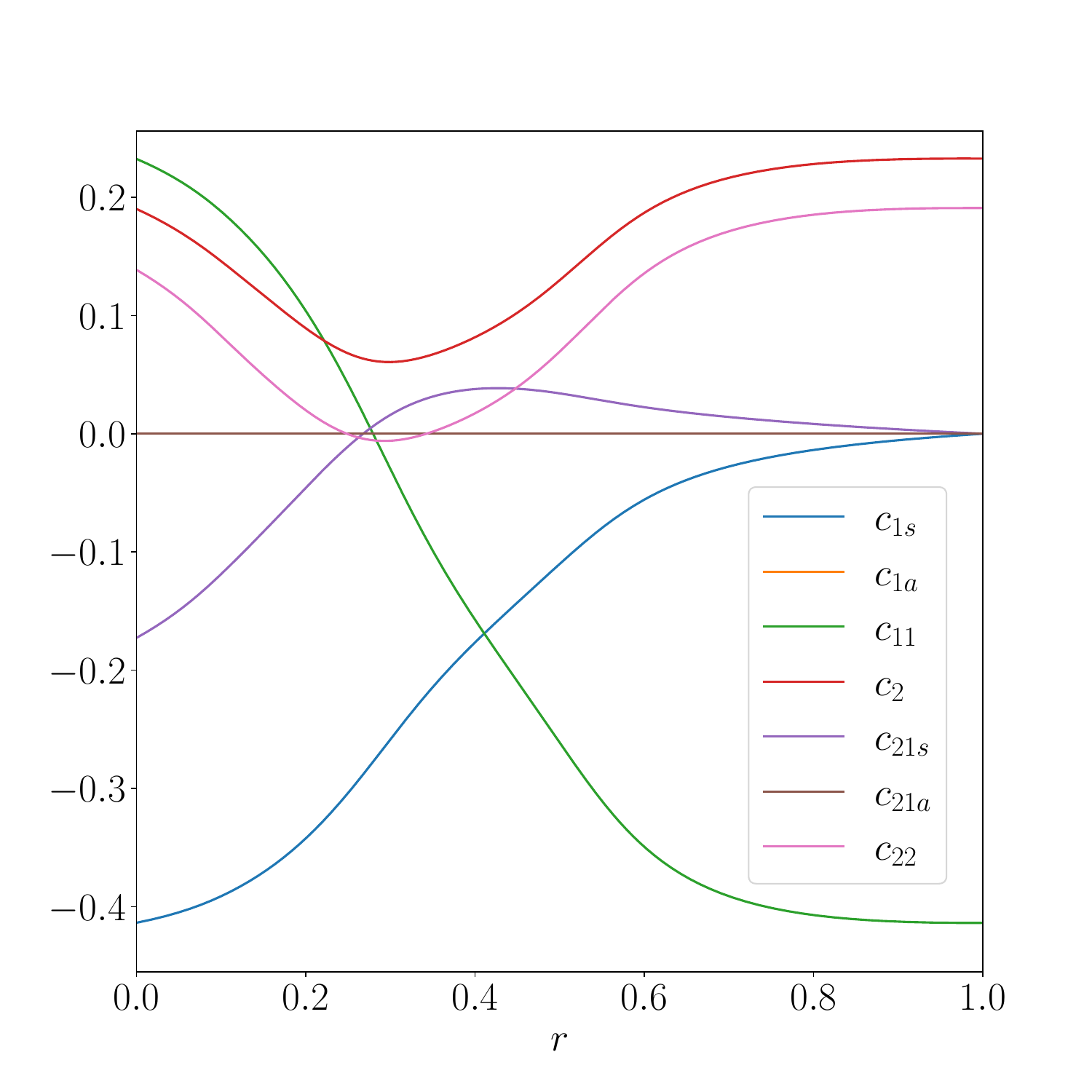}
  \caption{
    \label{fig:J1J2_T0_r}
    (Color online)
    Spin correlation functions as a function of $r=J_2/J_1$ at $T=0$.
    There is symmetry between the correlation functions at $r=0$
    and those at $r=1$.
    Specifically, $c_1$, $c_{11}$, and $c_2$ at $r=0$
    correspond to $c_{11}$, $c_2$, and $c_{22}$ at $r=1$,
    respectively.
    The results confirm this symmetry.
    Additionally, the nematic correlations, $c_{1a}$ and $c_{21a}$, are found to be zero
    within the numerical error.
  }
\end{figure}

Figure~\ref{fig:J1J2_T0_r} presents the results at temperature $T=0$.
The correlation functions are plotted as a function of the ratio $r=J_2/J_1$.
It is worth noting that there exists a symmetry between the cases $r=0$ and $r=1$,
as previously observed \cite{Siurakshina2001},
due to the exchange between $J_1$ and $J_2$.
Additionally, we observe that the nematic correlations $c_{1a}=a_{1a}/\alpha$
and $c_{21a}=a_{21a}/\alpha$ are both zero within the numerical error.
The formula derived in this study was obtained by employing the decoupling scheme
at the second-order equation of motion.
Within this approximation, the right hand side of Eq.~(\ref{eq:I}) diverges at $T=0$ when
the spin-wave gap associated with the long-range order vanishes.
Therefore, the self-consistent equations at $T=0$
do not directly allow us to discuss the presence of long-range order.
However, within this framework, we can investigate the short-range order.
Based on our analysis of the self-consistent equations,
we find no indication of short-range nematic correlation in the system.
%-------------------------------------------------------------------
%-------------------------------------------------------------------

\begin{figure}[htbp]
  \includegraphics[width=\linewidth]{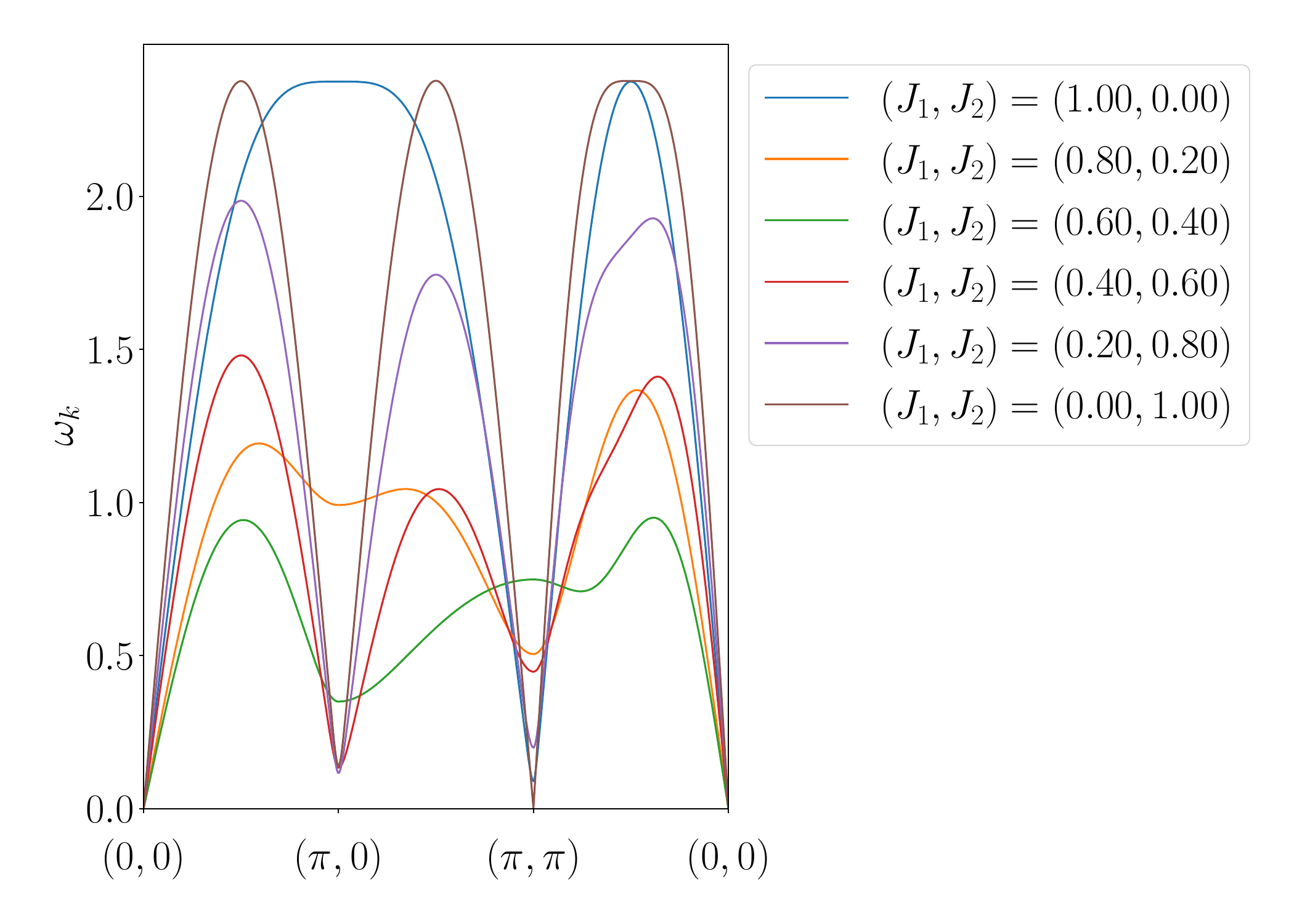}
  \caption{
    \label{fig:J1J2_disp}
    (Color online)    
    Spin-wave dispersion along the high-symmetric lines in the Brillouin zone
    for different values of $J_1$ and $J_2$.
    Here, the values of $J_1$, $J_2$, and $\omega_{\bm{k}}$
    are expressed in units of $J$, which is taken as the unit of energy.
  }
\end{figure}

The spin-wave dispersion at $T=0$ is shown in Fig.~\ref{fig:J1J2_disp}.
There is a systematic change in the energy dispersion
by varying the values of $J_1$ and $J_2$.
It is important to note that there is no need to assume any long-range order
to obtain the spin-wave dispersion.
However, a discrepancy arises when considering the gap at $(\pi,\pi)$ and $(\pi,0)$.
For the case when $J_1$ is finite and $J_2=0$, the gap at $(\pi,\pi)$ should vanish,
and similarly, for $J_1=0$ and $J_2$ is finite, the gap at $(\pi,0)$ should also vanish.
However, the decoupling approximation at the second-order equation of motion,
as mentioned earlier,
leads to a finite gap at these points, contradicting the expected behavior.
Employing the decoupling scheme at the third-order equation of motion
has the potential to yield improvements and could resolve
the discrepancy observed in the gap at $(\pi,\pi)$ and $(\pi,0)$.
Such investigations are left for future research.
One can also compute the energy dispersion at finite temperature
without long-range order but with short-range order.
The advantage of this formula is that it enables the computation
of dynamical quantities, which can be experimentally verified.
%-------------------------------------------------------------------
%-------------------------------------------------------------------

\begin{figure}[htbp]
  \includegraphics[width=0.8 \linewidth, angle=0]{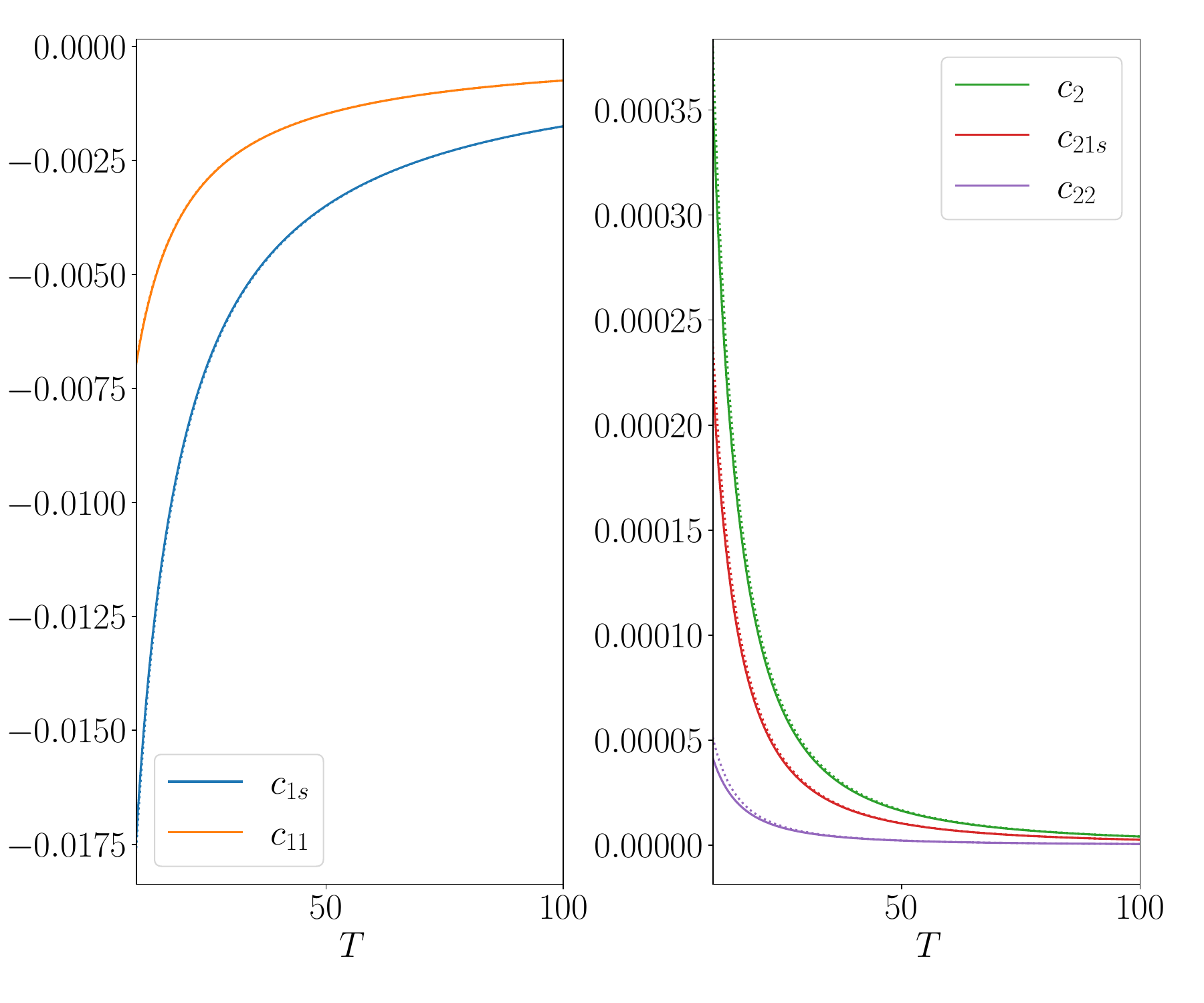}
  \caption{
    \label{fig:J1J2:highTexp}
    (Color online)
      Temperature dependence of the correlation functions (dotted lines)
      compared with the high-temperature expansion results,
      given by Eqs.~(\ref{eq:J1J2:highT:c1s}) to (\ref{eq:J1J2:highT:c22}) (solid lines),
      for $J_1=0.7$ and $J_2=0.3$.
      The temperature range is $10 \leq T \leq 100$.
  }
\end{figure}

  We now investigate the model at finite temperatures.
We expect the Green's function approach to become accurate at high temperatures
due to the short spin-spin correlation length.
  This is explicitly demonstrated for the one-dimensional system
  and the square lattice case above.
Therefore, in principle, we can utilize the high-temperature expansion result\cite{Hutak2022}
to set the initial values of the correlation functions, $c_{ij}$,
when solving the self-consistent equations.
  The high-temperature expansion results for the correlation functions,
  up to the second order in $\beta$, are given by:
  \bea
      {c_{1s}} &=&  - \frac{1}{4}\beta {J_1}
      - \frac{1}{{16}}{\beta ^2}J_1^2 + \frac{1}{8}{\beta ^2}{J_1}{J_2},
      \label{eq:J1J2:highT:c1s}
      \\
      {c_{11}} &=&  - \frac{1}{4}\beta {J_2} + \frac{1}{8}{\beta ^2}J_1^2,\\
      {c_2} &=& \frac{1}{{16}}{\beta ^2}J_1^2 + \frac{1}{8}{\beta ^2}J_2^2,\\
      {c_{21s}} &=& \frac{1}{8}{\beta ^2}{J_1}{J_2},\\
      {c_{22}} &=& \frac{1}{{16}}{\beta ^2}J_2^2.
      \label{eq:J1J2:highT:c22}      
      \eea
      Figure \ref{fig:J1J2:highTexp} presents the correlation functions
      compared with the results from high-temperature expansions.
      The concordance is excellent at elevated temperatures.
%-------------------------------------------------------------------
%-------------------------------------------------------------------

\begin{figure}[htbp]
  \includegraphics[width=1 \linewidth, angle=0]{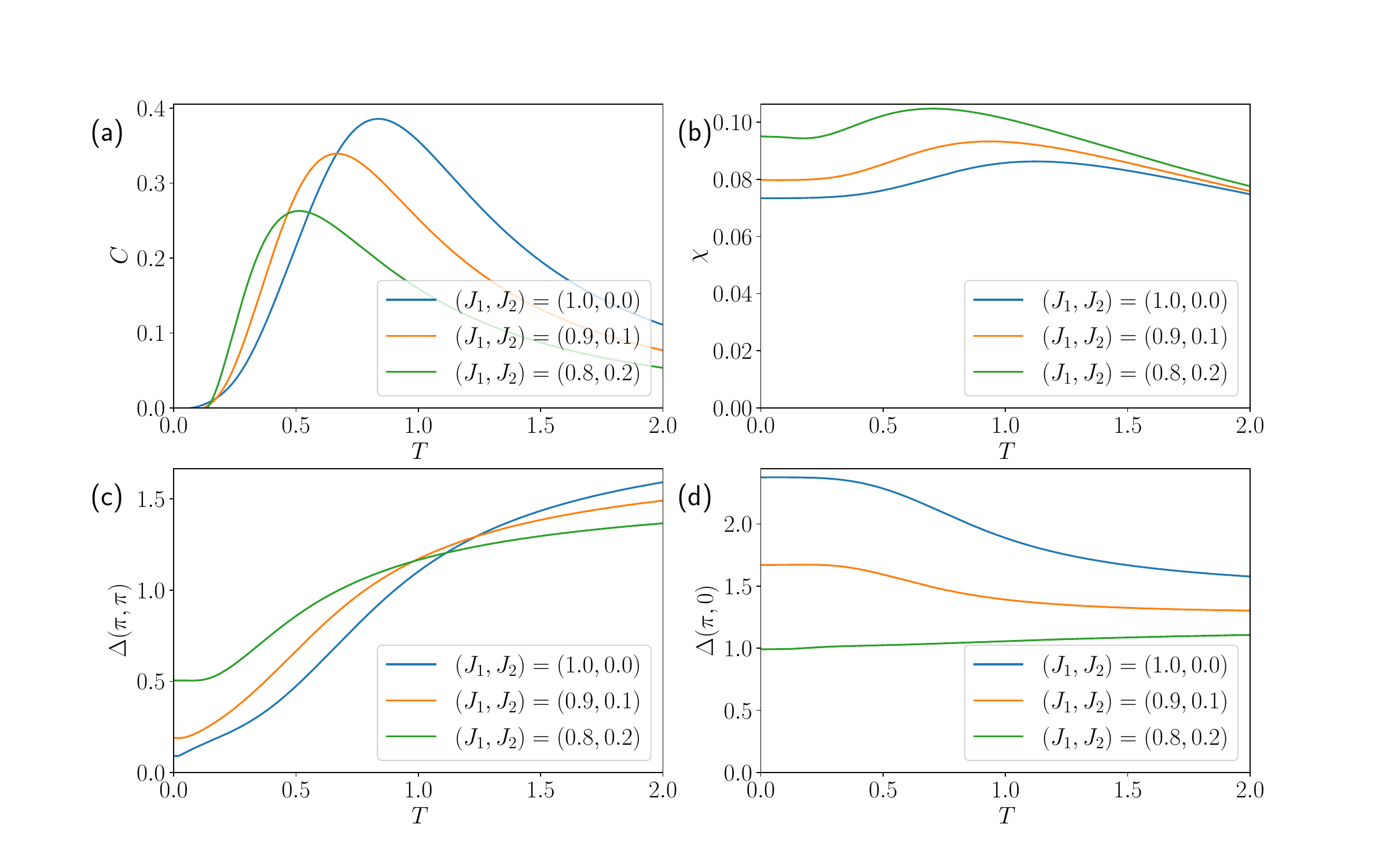}
  \caption{
    \label{fig:J1J2:large_J1}
    (Color online)
      Temperature dependencies for (a) the specific heat,
      (b) the spin susceptibility,
      (c) the gap at $(\pi,\pi)$,
      and (d) the gap at $(\pi,0)$
      for the cases of $(J_1,J_2)=(1.0,0.0),(0.9,0.1),(0.8,0.2)$.
  }
\end{figure}

\begin{figure}[htbp]
  \includegraphics[width=1 \linewidth, angle=0]{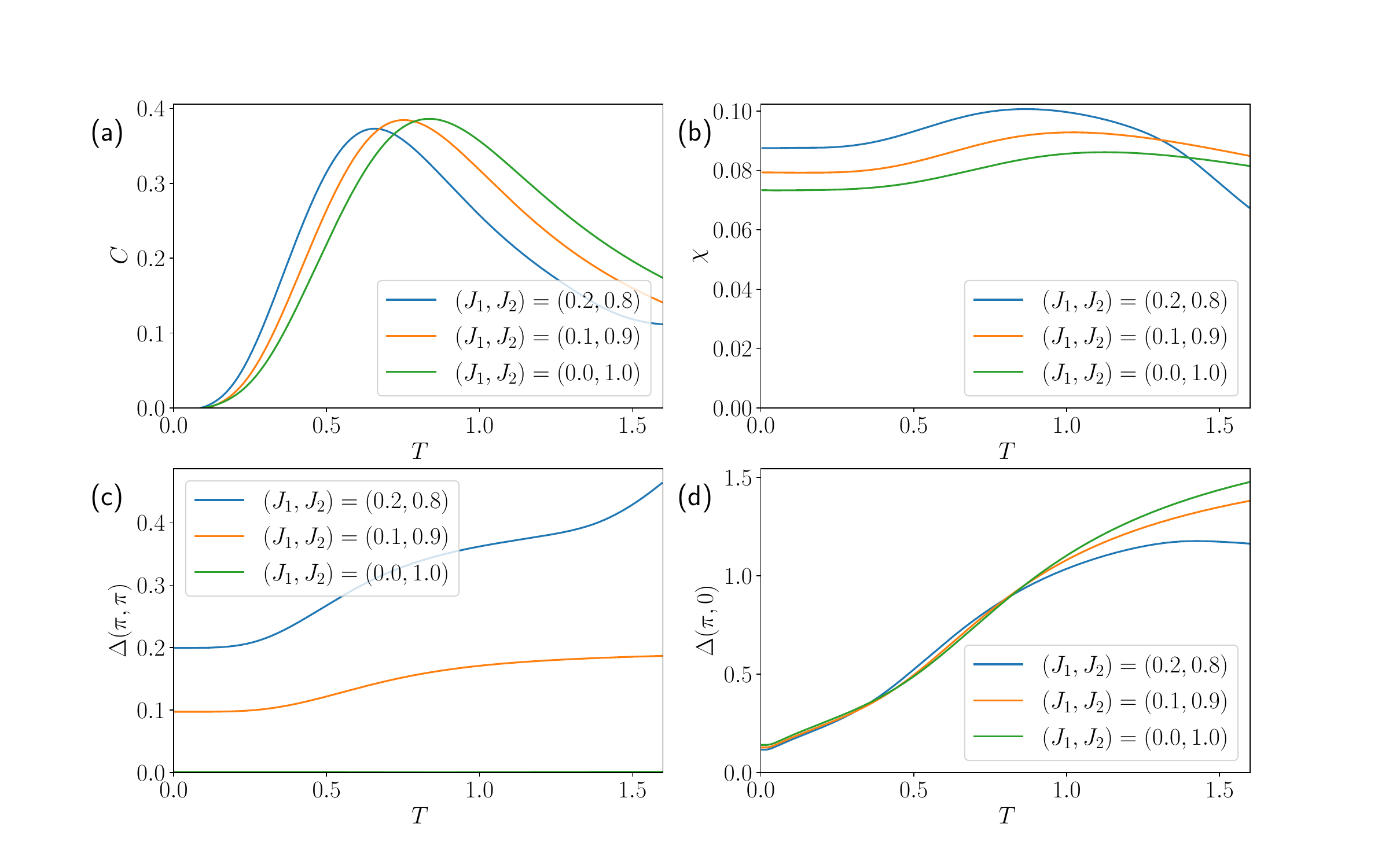}
  \caption{
    \label{fig:J1J2:large_J2}
    (Color online)
      Temperature dependencies for (a) the specific heat,
      (b) the spin susceptibility,
      (c) the gap at $(\pi,\pi)$,
      and (d) the gap at $(\pi,0)$ for the cases of
      $(J_1,J_2)=(0.2,0.8),(0.1,0.9),(0.0,1.0)$.
  }
\end{figure}

      In principle, one can perform a self-consistent calculation
      by initializing the correlation functions
      with the high-temperature expansion results
      and then progressively lowering the temperature.
      This method is particularly effective
      when the effects of frustration are weak.
      For the cases of $(J_1,J_2)=(1,0),(0.9,0.1),(0.8,0.2)$
      and $(J_1,J_2)=(0,1),(0.1,0.9),(0.2,0.8)$,
      we achieved satisfactory results.
      The former is shown in Fig.~\ref{fig:J1J2:large_J1}
      and the latter in Fig.~\ref{fig:J1J2:large_J2},
      where the temperature dependencies of the specific heat,
      spin susceptibility,
      and gaps at characteristic wave vectors are shown.
      Here, $\chi$ and $C$ are measured in units of
      ${\left( {g{\mu _{\rm{B}}}} \right)^2}/J$
      and $N{k_{\rm{B}}}J$,
      respectively, where $J$ represents the unit of energy. 
      The gaps are defined as
      $\Delta \left( {\pi ,\pi } \right) = {\omega _{\left( {\pi ,\pi } \right)}}$
      and
      $\Delta \left( {\pi ,0} \right) = {\omega _{\left( {\pi ,0} \right)}}$.
      In the case of small $J_2$, as shown in Fig.~\ref{fig:J1J2:large_J1},
      the peak in the specific heat shifts to lower temperatures
      with an increase in $J_2$,
      while simultaneously reducing the peak value.
      This shift is also observed in the spin susceptibility.
      This trend can be interpreted as the suppression of the antiferromagnetic
      correlation with the wave vector $(\pi,\pi)$
      due to the increase in $J_2$.
      This interpretation aligns with the behavior of $\Delta(\pi, \pi)$,
      which increases as $J_2$ increases,
      indicating that the antiferromagnetic correlation becomes
      more short-ranged.
      Meanwhile, $\Delta(\pi, 0)$ decreases with an increase in $J_2$,
      suggesting a slight enhancement
      of the antiferromagnetic correlation with the wave vector $(\pi,0)$.

      In the case of small $J_1$, as depicted in Fig.~\ref{fig:J1J2:large_J2},
      the peak in the specific heat shifts to lower temperatures
      with an increase in $J_1$.
      Unlike the case with small $J_2$,
      the peak value remains almost unchanged.
      This shift is also observed in the spin susceptibility,
      interpreted as the suppression of the antiferromagnetic correlation
      with the wave vector $(\pi,0)$ due to the increase in $J_1$.
      For the case of $(J_1,J_2)=(0,1)$, the gap $\Delta(\pi, \pi)$
      is identically zero.
      $\Delta(\pi, \pi)$ becomes finite and increases as $J_1$ increases.
      Meanwhile, $\Delta(\pi, 0)$ shows almost no change.
      This behavior suggests that the system is close to the antiferromagnetic
      long-range ordered state with the wave vector $(\pi,0)$.

%-------------------------------------------------------------------
%-------------------------------------------------------------------

%-------------------------------------------------------------------
%-------------------------------------------------------------------

\begin{figure}[htbp]
  \includegraphics[width=1 \linewidth, angle=0]{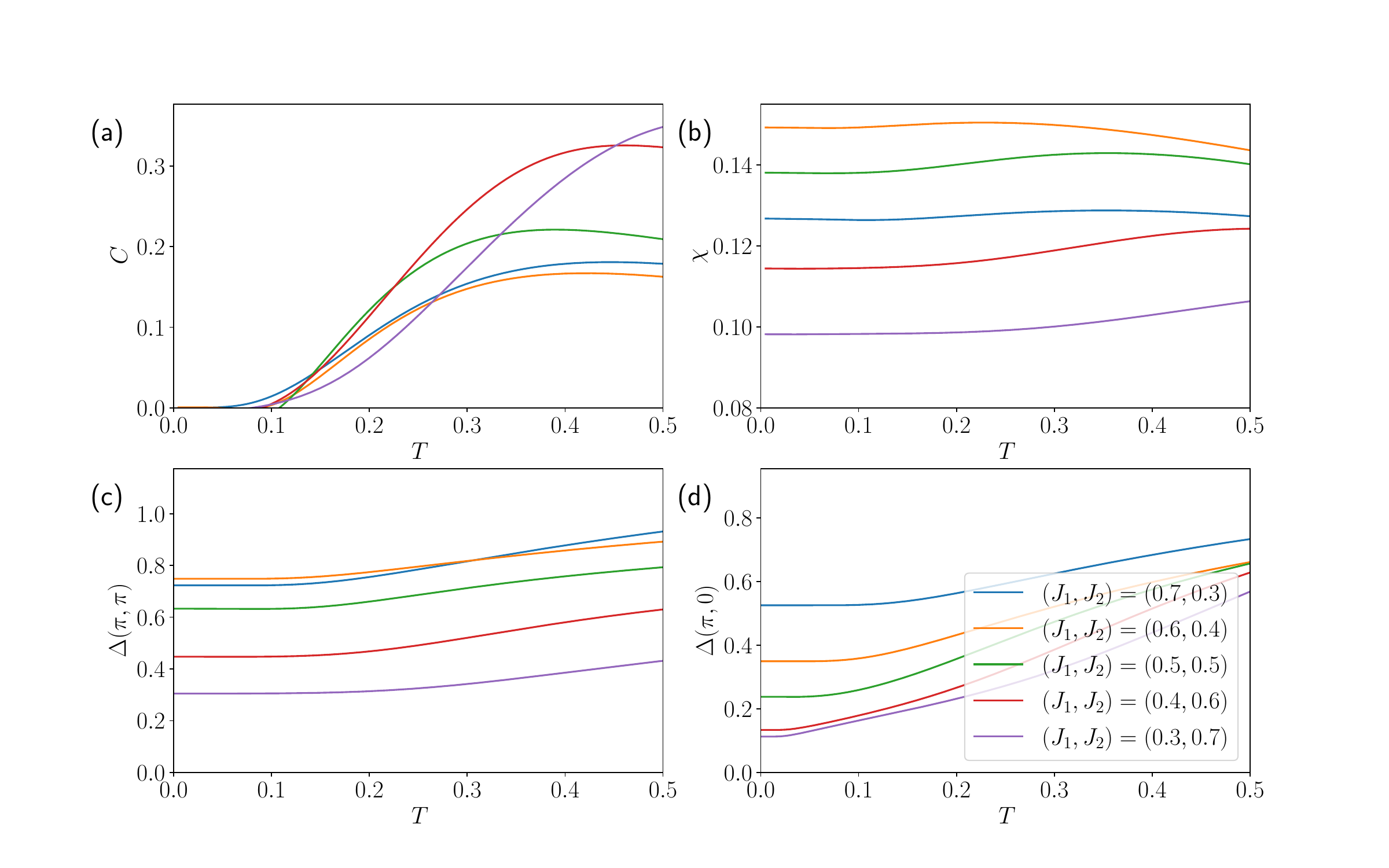}
  \caption{
    \label{fig:J1J2:frust}
    (Color online)
      Temperature dependencies for (a) the specific heat,
      (b) the spin susceptibility,
      (c) the gap at $(\pi,\pi)$,
      and (d) the gap at $(\pi,0)$ for the cases of
      $(J_1,J_2)=(0.7,0.3),(0.6,0.4),(0.5,0.5),(0.4,0.6),(0.3,0.7)$.
  }
\end{figure}

  Now we consider the cases with strong frustration.
  When the effects of frustration are pronounced,
  extremely precise numerical computations
  become necessary\cite{Hutak2022}.
  In fact the procedure of starting from high temperatures
  and lowering them failed
  for the cases of $(J_1,J_2)=(0.7,0.3),(0.6,0.4),(0.5,0.5),(0.4,0.6),(0.3,0.7)$.
  For these cases we take an alternative approach of starting
  from the zero-temperature results and then increase the temperature.
  The results are shown in Fig.~\ref{fig:J1J2:frust}.
  The variations in the specific heat and spin susceptibility
  are non-monotonic.
  Initially, the peak in the specific heat decreases
  and then increases as $J_2$ is raised.
  Similarly, the magnitude of the spin susceptibility first increases
  and then decreases with increasing $J_2$,
  indicating enhanced fluctuations around $(J_1,J_2) \sim (0.6,0.4)$.
  In contrast, the changes in the gaps are more uniform.
  Both $\Delta(\pi, \pi)$ and $\Delta(\pi, 0)$ decrease
  almost monotonically as $J_2$ increases.
  For the case of $(J_1,J_2) \sim (0.6,0.4)$,
  the magnitudes of the gaps suggest a lack of
  significant antiferromagnetic correlations
  associated with the interactions $J_1$ and $J_2$,
  indicating that the system is distant from the long-range ordered state.
  While the results seem plausible,
  they are not entirely convincing.
  Clearly, further numerical calculations
  that correlate with the high-temperature expansion results are necessary.
  Such studies are reserved for future research.

%-------------------------------------------------------------------
%-------------------------------------------------------------------

To obtain reliable results for the $J_1$-$J_2$ model,
there are two possible approaches.
One option is to perform a three-dimensional calculation
by considering a quasi-two-dimensional system.
Such a calculation was carried out\cite{Schmalfuss2006}
using several decoupling parameters and incorporating input from exact diagonalization
while introducing an approximate expression for the ground state energy.
If we introduce one decoupling parameter,
there is no need to consider inputs from other methods.
Another approach involves implementing the decoupling approximation
at the third-order equation of motion.
These aspects remain to be explored in future research.
\\

\section{Summary}
To summarize, we have presented
a generalized formulation of the Green's function method
that can be applied to diverse antiferromagnetic Heisenberg spin systems.
We have discussed
the hypercubic lattice and the $J_1$-$J_2$ model as specific applications.
We have derived the spin-wave dispersion formula
for the hypercubic lattice in any spatial dimension.
  We have shown that for $d=1$ and $d=2$, our formula yields accurate results
  at high temperatures, and for $d=3$, it predicts the transition temperature
  with greater precision than mean field theory,
  closely aligning with the quantum Monte Carlo results.
For the case of the $J_1$-$J_2$ model,
we computed the correlation functions at zero temperature
and spin-wave dispersion.
Although we have included the effect of nematic correlation,
there is no signature of the nematic order.
  At finite temperatures, we calculate the specific heat, spin susceptibility,
  and gaps at characteristic wave vectors.
  We achieve reasonable results when the effect of frustration is weak.
  However, we are unable to obtain results consistent
  with the high-temperature expansion when the effect of frustration is pronounced.

We note that the Green's function approach can be
generalized to incorporate other interactions and higher spin values.
The key features of this approach are that it does not need to assume specific correlations
and enables the investigation of finite temperature and dynamical properties.

Exploring higher-order approximations in the decoupling scheme is promising,
and it may provide more accurate results and a better understanding
of the physical behavior of the spin system.
Such investigation can open up new insights into the spin system
and pave the way for more accurate and comprehensive theoretical descriptions.
\bibliography{../../../../refs/lib}
\end{document}